\documentclass[runningheads]{llncs}

\usepackage{eccv}

\usepackage{eccvabbrv}

\usepackage{graphicx}
\usepackage{booktabs}

\usepackage[accsupp]{axessibility}  

\usepackage[pagebackref,breaklinks,colorlinks,citecolor=eccvblue]{hyperref}

\usepackage{orcidlink}

\begin{document}


\title{Event-Stream Super Resolution using Sigma Delta Neural Network} 

\titlerunning{Event Stream Super Resolution}


\author{Waseem Shariff\inst{*1, 2}\orcidlink{0000-0001-7298-9389} \and
Joe Lemley\inst{2}\orcidlink{0000-0002-0595-2313} \and
Peter Corcoran\inst{1}\orcidlink{0000-0003-1670-4793}}

\authorrunning{W. Shariff et al.}


\institute{
    C3I Imaging Laboratory, University of Galway, Ireland \\
    \email{\{w.shariff1, peter.corcoran\}@universityofgalway.ie}\\
    \url{https://www.universityofgalway.ie/c3i/}\\ \and
    FotoNation-Tobii, Galway, Ireland\\
    \email{\{joeseph.lemley\}@tobii.com}
}

\maketitle
\begin{abstract}

This study introduces a novel approach to enhance the spatial-temporal resolution of time-event pixels based on luminance changes captured by event cameras. These cameras present unique challenges due to their low resolution and the sparse, asynchronous nature of the data they collect. Current event super-resolution algorithms are not fully optimized for the distinct data structure produced by event cameras, resulting in inefficiencies in capturing the full dynamism and detail of visual scenes with improved computational complexity. To bridge this gap, our research proposes a method that integrates binary spikes with Sigma Delta Neural Networks (SDNNs), leveraging spatiotemporal constraint learning mechanism designed to simultaneously learn the spatial and temporal distributions of the event stream. The proposed network is evaluated using widely recognized benchmark datasets, including N-MNIST, CIFAR10-DVS, ASL-DVS, and Event-NFS. A comprehensive evaluation framework is employed, assessing both the accuracy, through root mean square error (RMSE), and the computational efficiency of our model. The findings  demonstrate significant improvements over existing state-of-the-art methods, specifically, the proposed method outperforms state-of-the-art performance in computational efficiency, achieving a 17.04-fold improvement in event sparsity and a 32.28-fold increase in synaptic operation efficiency over traditional artificial neural networks, alongside a two-fold better performance over spiking neural networks. 
  \keywords{Event-based Vision \and Super Resolution \and Sigma-Delta Neural Networks \and Computational Complexity}
\end{abstract}

\section{Introduction}
\label{sec:intro}

Neuromorphic event vision sensors have brought new sensing capabilities to computer vision. Unlike traditional cameras that capture images at regular intervals, event cameras (EC) detect changes in brightness in the scene as they happen, offering a unique way of sensing without seeing \cite{shariff}. Among these, the Prophesee Gen 4 cameras offer high-definition event-streams. However, sensor size remains critical for cost-sensitive applications, such as driver monitoring systems (DMS) and AR/VR technologies \cite{shariff}. In environments where space and power are constrained, compact designs like the GenX320 sensors \cite{genx} and the Dynamic Vision Sensor (DVS) by iniVation \cite{dvs} prove to be practical solutions.


However, these event cameras, while capable of operating at higher frame rates than conventional cameras, are typically limited to a spatial resolution of 320x320. This limitation restricts their use in applications requiring accurate detection and analysis, despite their potential advantages over RGB and NIR imaging systems. This study investigates the possibility of algorithmically enhancing the spatial-temporal resolution of event cameras using their high temporal resolution. The objective is to develop computationally efficient algorithms capable of interpolating high-resolution (HR) event streams from low-resolution (LR) data, thus addressing the resolution limitations of these sensors. Such improvements could expand the use of event cameras across various disciplines by leveraging their small size and other unique benefits.

\begin{figure}[h]
  \centering
   \includegraphics[width=1\linewidth]{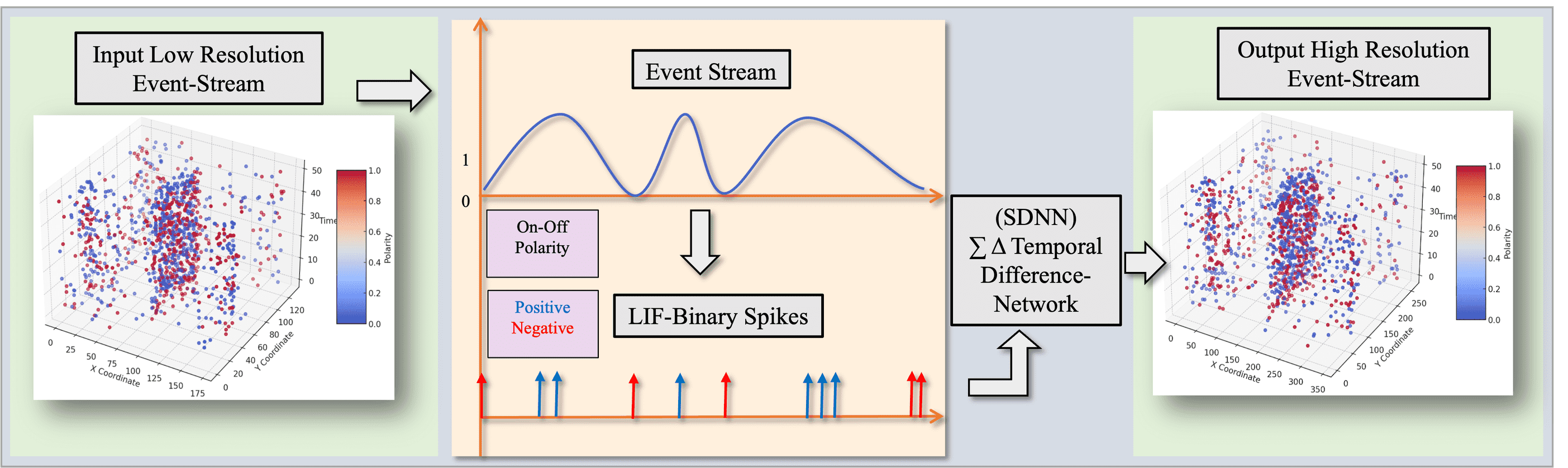}
   \caption{Overview of the proposed system, combining binary spike inputs with SDNNs' temporal difference mechanism to further enhancing event stream resolution.}
   \label{fig:overview}
\end{figure}

 Most applications process event-stream data by converting it into temporal frames or clusters, which can lead to loss of some amount of information. The proposed approach processes the asynchronous event stream directly, aiming to generate high-resolution spatiotemporal event streams and preserve spatial-temporal accuracy. Spiking Neural Networks (SNNs) \cite{snn} and Sigma Delta Neural Networks (SDNNs) \cite{sekikawa2021learning} represent two promising paradigms. SNNs are notable for their temporal precision and energy efficiency, as they process binary spikes asynchronously, significantly reducing computational and communication loads. SDNNs, on the other hand, excel in processing input signals with high precision and noise immunity, leveraging sigma-delta modulation to efficiently convert continuous signals into discrete forms, thus promoting computational efficiency and sparsity. \cref{fig:overview} provides an overview of the proposed system. This research strives to merge the strengths of SNNs (binary spikes as input) and SDNNs—utilizing the temporal precision of spikes and the accuracy of sigma-delta modulation—to improve the quality of low-resolution event streams.

Nevertheless, the exploration of the event stream super resolution is relatively limited, with only a few attempts, such as \cite{li}'s method utilizing sparse signal representation and non-homogeneous Poisson processes to predict event timestamps to further produce high-resolution event streams. Further \cite{snn} proposed utilizing the spiking networks while significantly reducing the error between spatial and temporal resolution. However, a common shortfall has been their struggle to computationally preserve these signal changes. Moreover, these existing techniques often exhibit limited robustness to noise, significantly constraining their effectiveness and applicability. This underscores the necessity for developing methods that are both more robust and computational efficient. In light of this challenge, the following are some of contributions of this paper.

\begin{itemize}
    \item This research introduces a novel end-to-end event stream super-resolution technique derived from use of sigma-delta neural network (SDNN).
    \item  The proposed methodology not only aims to improve spatial and temporal resolution but also focuses on computational efficiency.
    \item  Extensive validation on four benchmark event datasets—N-MNIST, CIFAR10-DVS, ASL-DVS and the real-event super-resolution Event-NFS dataset.
\end{itemize}

\section{Background}

This section provides literature review of the development of sigma-delta neural networks over a period and explains their suitability as a computationally efficient solution. It will also cover the background of both event-frame based and event-stream super resolution methods.
\subsection{Sigma-Delta Quantization}
The exploration of Sigma Delta Neural Networks (SDNNs) has witnessed significant advancements, each building upon the insights and methodologies introduced in earlier studies. Authors in \cite{ref2}, proposed the concept of an Adapting Spiking Neural Network (ASNN) grounded in adaptive spiking neurons and Asynchronous Pulsed Sigma-Delta coding. This approach efficiently encodes information in spike trains, optimizing firing rates through homeostatic mechanisms. Yoon et al. \cite{ref3} extends this concept by investigating how Leaky Integrate-and-Fire (LIF) and simplified Spike Response Model (SRM) neurons employ a similar Asynchronous Pulse Sigma-Delta Modulation (APSDM) scheme to encode continuous-time signals into spikes. The practical application of such encoding methods is demonstrated by \cite{ref8}, who apply APSDM to event-based sensors like the Dynamic-Vision Sensor, showcasing its potential for efficiently processing changes in analog pixel voltages.

Advancements in SDNNs continue with the work of \cite{ref4}, who propose spiking deep networks as an energy-efficient alternative. Their study focuses on minimizing performance loss during the conversion process from analog neural networks, achieving high accuracy with reduced training time \cite{ref4}. Courbariaux et al. \cite{ref5} introduced Binarized Neural Networks (BNNs), demonstrating their effectiveness in reducing memory size and arithmetic operations, achieving state-of-the-art results across various datasets. 

The trajectory of research in SDNNs continues to unfold as shown by \cite{sdnn}. \cite{ref9} introduces a backpropagation model for spiking networks using time-agnostic spiking neurons, building on principles from \cite{ref2} and \cite{ref10}. O’Connor et al. \cite{sdnn} further explore quantized temporal differences for efficient event-based sensor processing. These studies highlight the interdisciplinary nature of SDNN research, demonstrating how advancements in spiking neural networks, temporal encoding, and event-based sensors synergistically enhance our understanding of efficient neural computation. Additionally, Sekikawa et al. \cite{sekikawa2021learning} proposed a difference-driven SDNN framework for event processing, using histograms of fixed event counts to show performance. The ongoing refinement of these concepts promises more energy-efficient and high-performance neural network architectures.


\subsection{Event based Super Resolution}

To address event-based super resolution, researchers have developed various algorithms and networks for event-based super-resolution. Some novel techniques directly reconstruct high-resolution images from event streams. \cite{ref11} introduced an novel end-to-end network for high-resolution and HDR image reconstruction. \cite{ref12} developed EventSR, an unsupervised pipeline with superior image quality. Authors in  \cite{ref13, ref14, ref15, ref16, ref17, neurozoom} explored event-based denoising events, event-zoom, video super-resolution and reconstruction of event stream, achieving state-of-the-art results. Other works focused on spatial-temporal super-resolution, \cite{ref18} used spatial-temporal interpolation, \cite{ref19} revitalized spike-based reconstruction of super-resolution streams, and \cite{ref20} employed a recurrent neural network to boost the super resolved event data.


Two recent works have explored the resolution enhancement of 'event-streams'. In the research conducted by \cite{li}, the focus lies in super-resolution (SR) for spatiotemporal event-stream images captured by frame-free dynamic vision sensors (DVSs). The authors introduce a two-stage scheme that utilizes a non-homogeneous Poisson point process to model event sequences. In the initial stage, the event count for each pixel in the high-resolution (HR) image is determined through sparse signal representation, while the second stage involves generating the event sequence using a thinning-based event sampling algorithm. Evaluation of the method involves obtaining HR event-stream images from DVS recordings, demonstrating improved spatial texture detail and well-matched temporal properties compared to low-resolution (LR) images. On the other hand, \cite{snn} present a comprehensive framework for super-resolving event streams using a spiking neural network. This research adopted their approach that incorporates a spatiotemporal constraint learning mechanism to simultaneously learn the spatial and temporal distributions of event streams. Validation on large-scale datasets showcases state-of-the-art performance in object classification and image reconstruction. Both studies contribute to advancing super-resolution techniques for event-based vision, addressing the unique challenges posed by spatiotemporal event-stream data and highlighting their applicability in various scenarios. However, their effectiveness is constrained by limitations in robustness to noise and computational efficiency.

\section{Methodology}
The proposed network adopts SDNN with a spiking synapses to further generate super resolved event stream. This section briefly outlines the theoretical formulation of SDNN and explains briefly the distinctions between Artificial Neural Networks (ANNs), Spiking Neural Networks (SNNs), and SDNNs.

\subsection{Sigma Delta Neural Network}

This work is inspired from Sigma-Delta network methodology \cite{sdnn}, which introduces an innovative approach to processing neural information by exploiting the temporal dynamics inherent in sequential data. Central to the SDNN formulation are two operations: Temporal Difference ($\Delta T$) and Temporal Integration ($\Sigma T$), which together facilitate the efficient handling of spatio-temporally redundant data. The Temporal Difference operation, defined by $y_t = x_t - x_{t-1}$, captures the change in input or activation from one time step to the next, emphasizing the network's focus on variations in the data. This operation is complemented by the Temporal Integration operation, $Y_t = Y_{t-1} + y_t$, which accumulates these changes over time, thereby integrating information across temporal dimensions \cite{sdnn}. 

To seamlessly integrate these operations within the network, the forward function of the network is redefined to include $\Delta T$ and $\Sigma T$ operations between each layer, transforming the network's processing approach without altering its fundamental function. The network's architecture is thus adapted to process inputs in terms of their temporal differences, leveraging the temporal redundancy in datasets such as event streams to reduce computational load. For linear transformations within the network, represented by $w(x)$, and nonlinear functions, denoted as $h(x)$, the temporal operations are strategically applied to ensure that $\Delta T(w(\Sigma T(x))) = w(x)$, demonstrating the commutative property of these operations with linear transformations.

Moreover, the discretization of the outputs from the Temporal Difference modules into sparse representations plays a crucial role in enhancing computational efficiency. This step is articulated through the equation $y_t = \text{round}(x_t - x_{t-1})$, where $\text{round}(\cdot)$ denotes the operation of rounding to the nearest discrete value. Through this discretization process, the SDNN effectively transforms continuous data into a form that is more amenable to sparse processing, significantly reducing the computational demands. Algorithm 1, provided in the supplementary material, outlines the working of SDNN. 

\begin{figure}[h]
  \centering
   \includegraphics[width=0.9\linewidth]{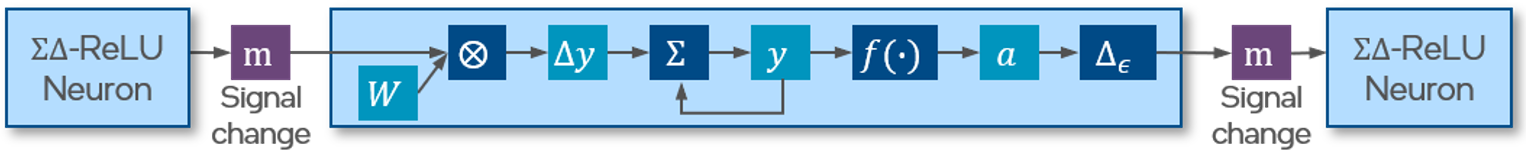}

   \caption{A sigma-delta neural network processing event-signals dynamically \cite{lava}.}
   \label{fig:sdnn}
\end{figure}

 \cref{fig:sdnn} provides a schematic representation of a neuron within a Sigma-Delta Neural Network (SDNN), which incorporates specialized operations for processing signals. It begins with the neuron labeled as $\Sigma\Delta$-ReLU Neuron, suggesting the utilization of a ReLU activation function in conjunction with sigma-delta modulation techniques. The signal processing flow initiates with a signal change denoted by 'm',  indicating the memory component of the neuron that retains the previous input's value. This leads to the Temporal Difference operation ($\Delta y$), which computes the difference between the current and the previous input signals, thus highlighting the network's emphasis on temporal changes. Following this, the Temporal Integration operation ($\Sigma$) aggregates these differences over successive time steps, effectively summing the dynamic changes of the input signal. The integrated signal, now represented by 'y', is subsequently passed through a non-linear function, indicated by $f(\cdot)$, potentially a ReLU function, to introduce non-linearity to the neuron's output. A subsequent Temporal Difference operation ($\Delta \epsilon$) measures the change post-activation function. At the output phase, the signal change 'm' suggests that the neuron is preparing the processed signal for delivery to the subsequent network layer or as the network's final output. This depiction of an SDNN neuron underscores the architecture's efficiency in processing sequential data by reducing redundancy and prioritizing the temporal dynamics of data, thereby concentrating computational resources on the most significant temporal changes.



\subsection{Event Representation}

In this study, events are encoded into binary spikes utilizing the Leaky Integrate-and-Fire (LIF) neuron model \cite{lava} before their integration into the network architecture. The following equations describe a leaky integrator model:

    \begin{equation}
        y[t] = (1 - \alpha)y[t-1] + x[t]
        \label{eq:1}
    \end{equation}
    
     \cref{eq:1}, \emph{Leaky Integration} equation models the temporal dynamics of a neuron or unit in a network. Here, \(y[t]\) represents the state of the neuron at time \(t\), \(x[t]\) is the external input to the neuron at time \(t\), and \(\alpha\) is a decay factor. The decay factor determines how much of the previous state \(y[t-1]\) is retained, with the term \((1 - \alpha)\) scaling down the previous state before adding the current input \(x[t]\), thereby updating the neuron's state.

    \begin{equation}
        s[t] = \begin{cases} 
        1, & \text{if } y[t] \geq \vartheta \\
        0, & \text{otherwise}
        \end{cases}
        \label{eq:2}
    \end{equation}
    
    \cref{eq:2} suggests, \emph{Thresholding and Spiking}. This introduces a spiking mechanism where \(s[t]\) is a binary variable that indicates whether the neuron spikes at time \(t\). A spike occurs (\(s[t] = 1\)) if the neuron's state \(y[t]\) exceeds a certain threshold \(\vartheta\). If the state is below the threshold, the neuron does not spike (\(s[t] = 0\)).

    \begin{equation}
        y[t] = y[t](1-s[t])
        \label{eq:3}
    \end{equation}
    
    \cref{eq:3} suggests, \emph{State Update with Spike-Dependent Reset.} After determining the spiking condition, the neuron's state is updated. If a spike occurs (\(s[t] = 1\)), the state of the neuron is reset to 0, mimicking the biological refractory period following a spike. This reset is modeled by multiplying the state \(y[t]\) by \(0\), effectively resetting it. If the neuron does not spike (\(s[t] = 0\)), its state remains unchanged.This model, known as a leaky integrate-and-fire (LIF) neuron, captures both the gradual integration of inputs over time and the discrete spiking behavior when inputs exceed a certain threshold. It aims to mimic the temporal dynamics and computational properties of biological neural systems.

\subsection{Network Architecture}
\label{sec:network}

\begin{figure}[h]
  \centering
   \includegraphics[width=1\linewidth]{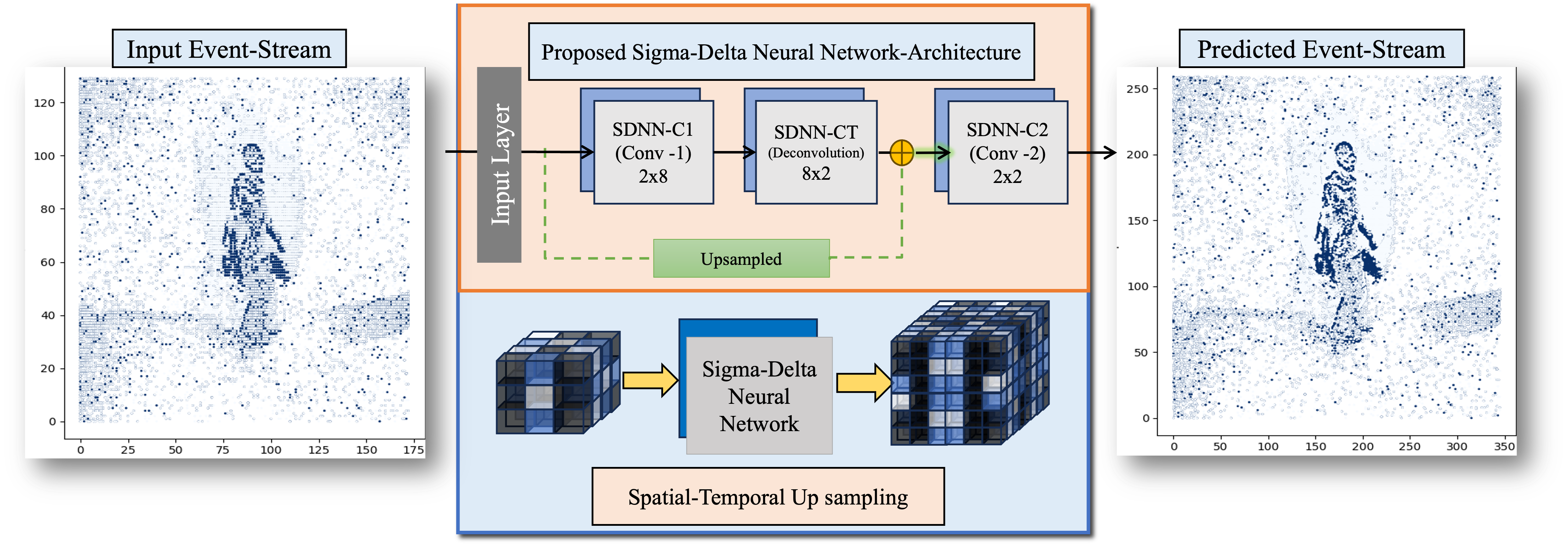}

   \caption{Sigma-Delta Neural Network architecture for enhancing event-stream data: The input event-stream (left) is processed through the proposed neural network, which includes convolution and deconvolution layers, resulting in a predicted event-stream (right) with improved spatial and temporal resolution.}
   \label{fig:network}
\end{figure}

Our network architecture is based on a convolutional SDNN with two convolution layers and one deconvolution layer, as shown in \cref{fig:network}. The network consists of three layers: \emph{sdnn\_c1}, \emph{sdnn\_ct}, and \emph{sdnn\_c2}.

The \emph{input-layer} includes mean-only batch normalization, sigma, and delta operations. The \emph{sdnn\_c1} module is a convolutional layer with 2 input channels and 8 output channels, a kernel size of (5, 5, 1), a stride of (1, 1, 1), and zero padding of (2, 2, 0). The \emph{sdnn\_ct} module is a transposed convolutional layer with 8 input channels and 2 output channels, a kernel size of (2, 2, 1), and a stride of (2, 2, 1). It includes a dropout operation and upsampled interpolation before combining with \emph{sdnn\_c2}. The \emph{sdnn\_c2} module has a single convolutional layer with 2 input channels and 2 output channels, using a 1x1 kernel and a stride of 1. Neuron processing in \emph{sdnn\_ct} and \emph{sdnn\_c2} involves a dropout factor of 0.1.

The architecture uses a convolutional layer after a deconvolutional layer to refine features and capture details. Deconvolution upsamples feature maps, and the subsequent convolutional layer refines and consolidates higher-resolution information, preserving local patterns that might be lost during upsampling. This approach balances capturing global context and preserving local details, helping in learning complex hierarchical features.

The SDNN also includes a parameter dictionary for sigma-delta neurons, with a delta unit threshold of 1, a surrogate gradient relaxation parameter (\emph{tau\_grad}) of 0.05, a surrogate gradient scale parameter (\emph{scale\_grad}) of 2, and a trainable threshold with shared parameters across the layer. The activation function for the neurons is Rectified Linear Unit (ReLU). The interaction between these parameters and the convolutional and deconvolutional layers allows the network to learn complex features, providing adaptability crucial for capturing spatiotemporal patterns during upsampling.

\subsection{Spatio-Temporal Learning}

The output HR event stream is expected to mirror the LR input's characteristics across both spatial and temporal dimensions. The spatial and temporal learning is adopted from \cite{li} and \cite{snn}, since these proved to be a more effective learning approach. In the temporal learning domain, the focus revolves around comprehending the dynamic evolution of events by minimizing the disparity between predicted and actual occurrences at each pixel across various timestamps. This involves formulating the temporal loss ($\mathcal{L}^T$) through the integration of mean squared differences over time, expressed as:
\begin{equation}
{Loss}^{Temporal}= \frac{1}{2} \sum_{i} \int_0^T \left(\mathcal{E}_i(t) - \widehat{\mathcal{E}}_i(t)\right)^2 \,dt
\label{eq:4}
\end{equation}

\cref{eq:4}, $\mathcal{E}_i(t)$ and $\widehat{\mathcal{E}}_i(t)$ signify the predicted and actual event occurrences, respectively, at the spatial coordinate $i$. The summation of triggered output events within a designated time interval [$T_0, T_1$] produces the Peristimulus Time Histogram (PSTH), a representative snapshot of spatial distribution. The learning of spatial characteristics involves minimizing the mismatch between the PSTH of the generated super resolved event stream and the ground high resolution event stream, encapsulated in the spatial loss ($\mathcal{L}^{Spatial}$):

\begin{equation}
{Loss}^{Spatial} = \frac{1}{2} \sum_{i} \left(\Phi_i - \widehat{\Phi}_i\right)^2
\label{eq:5}
\end{equation}

\cref{eq:5}, $\Phi_i$ and $\widehat{\Phi}_i$ denote the PSTH of the predicted and actual event streams, respectively, at the pixel with coordinate $i$. Practically, the time bin's length is manually defined as 50 milliseconds, devoid of any overlap. To combine temporal and spatial insights seamlessly, the overarching loss function ($\mathcal{L}$) is formulated as a fusion of temporal ($\mathcal{L}^{Temporal}$) and spatial ($\mathcal{L}^{Spatial}$) losses. The coefficients $\alpha$ and $\beta$ serve as hyperparameters, allowing a substantial adjustment of the influence of each component \cite{snn}:
\begin{equation}
\mathcal{L} = \alpha \cdot \mathcal{L}^{Temporal}+ \beta \cdot \mathcal{L}^{Spatial}
\label{eq:6}
\end{equation}

The overall approach, proved to be a more effective strategy, ensures a dynamic and versatile learning process, accounting for both temporal and spatial variations, with the flexibility to fine-tune their respective impacts through the introduced hyperparameters.

\section{Experiments and Results}
\subsection{Dataset}
This research utilized a dataset prepared by \cite{snn}. The ground truth for high-resolution (HR) event streams was established using the original DVS recordings, down-sampled by merging events within each 2×2 kernel with a stride of 2, producing low-resolution (LR) event streams. Evaluation was conducted on three publicly available datasets: N-MNIST \cite{nmnsit}, CIFAR10-DVS \cite{cifar}, and ASL-DVS \cite{asl}.
The N-MNIST dataset consists of 60,000 training samples and 10,000 test samples derived from the MNIST dataset using a moving ATIS sensor. The CIFAR10-DVS dataset includes dynamic images from the CIFAR10 dataset presented to a fixed DVS128 sensor, with 8,500 training samples and 1,500 test samples. The ASL-DVS dataset comprises sign language data, totaling 100,800 samples, with 75\% allocated for training and validation, and 25\% for testing. The approach by Siqi et al. \cite{snn-git} involved validation and testing on the same data, potentially contaminating their results and compromising performance analysis. In contrast, our study adopts a more robust methodology by first training and validating the network on the training data and then conducting testing exclusively on the designated unseen testing data.
\subsection{Implementation Details}
The proposed model undergoes training for 25 epochs, employing a batch size of 32 on 2 x Nvidia 2080 Ti GPUs. We used the Adam optimizer with an initial learning rate of 0.1, which was then reduced by a factor of 0.1 after every 6 epochs. This implementation utilizes the publicly available Lava-DL library \cite{lava}. Proposed SDNN is designed to detect both spatial and temporal changes in input while operating as a continuous-time system, requires discretization for GPU simulation. To achieve this, we set a simulation step size in milliseconds, with specific simulation duration same as \cite{snn} of 350, 200, and 1500 milliseconds for N-MNIST, ASLDVS, and CIFAR10-DVS Dataset respectively, accounting for variations in sample duration. 

\subsection{Quantitative Results}
\cref{tab:results} shows quantitative results of proposed and literature models measured by the root mean squared error (RMSE) across three datasets, N-MNIST, CIFAR10-DVS, and ASL-DVS. \cite{li} presented a baseline model that demonstrated competitive performance, achieving RMSE values of 0.757, 0.404, and 0.550 for N-MNIST, CIFAR10-DVS, and ASL-DVS, respectively. In contrast, the traditional Spiking Neural Network \cite{snn} exhibited better loss performance, indicating accurate super resolution across all three datasets. The proposed SDNN introduced an alternative architecture that, on average, outperformed the SNN \cite{snn}. 

\begin{table}[h!]
    \caption{Quantitative Evaluation: Comparison of Root Mean Square Error (RMSE) Loss between the Proposed SDNN Model and Existing Methods as reported in \cite{li} and \cite{snn} for Event Stream Super Resolution}
    \centering
    \begin{tabular}{lccc}
        \toprule
        \textbf{Model} & \textbf{N-MNIST} & \textbf{CIFAR10-DVS} & \textbf{ASL-DVS} \\
        \midrule
        \textbf{li et al.} \cite{li} & 0.757 & 0.404 & 0.550 \\
        \textbf{Siqi et al.} \cite{snn} & 0.272 & 0.179 & 0.229 \\
        \midrule
        \textbf{Proposed SDNN} & \textbf{0.258} & \textbf{0.174} & \textbf{0.214} \\
        \bottomrule
    \end{tabular}
    \label{tab:results}
\end{table}

\begin{figure}[h]
	\centering
	\begin{subfigure}{0.17\linewidth}
		\includegraphics[width=\linewidth]{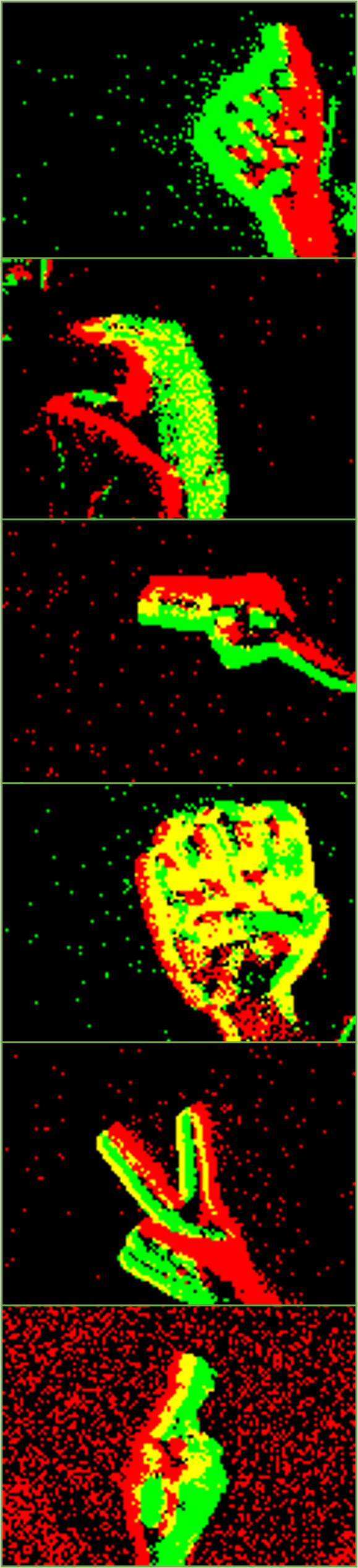}
	 \caption{Low- Resolution Input}\label{fig:5a}
	\end{subfigure}
	\begin{subfigure}{0.17\linewidth}
		\includegraphics[width=\linewidth]{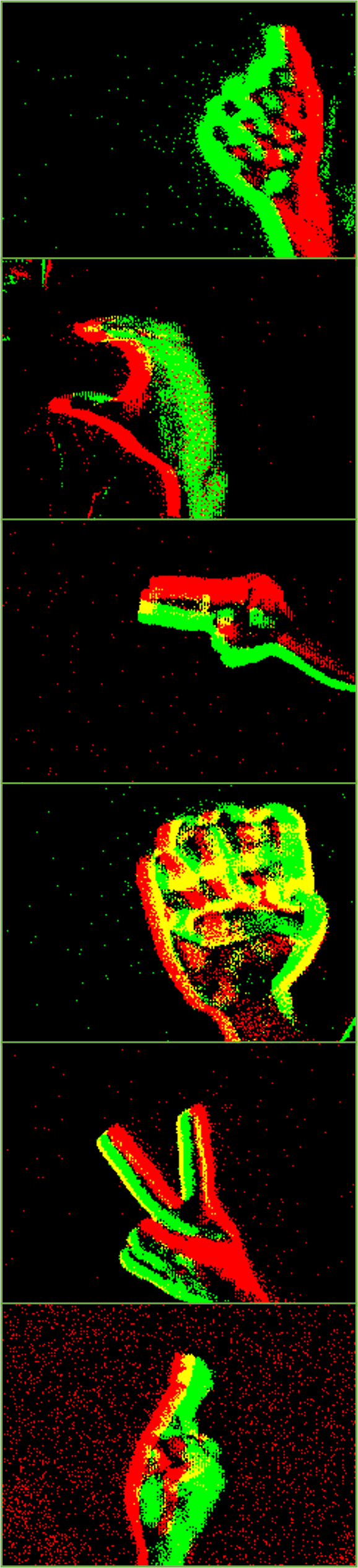}
		 \caption{Ground Truth-HR}\label{fig:5b}
	\end{subfigure}
	\begin{subfigure}{0.17\linewidth}
		\includegraphics[width=\linewidth]{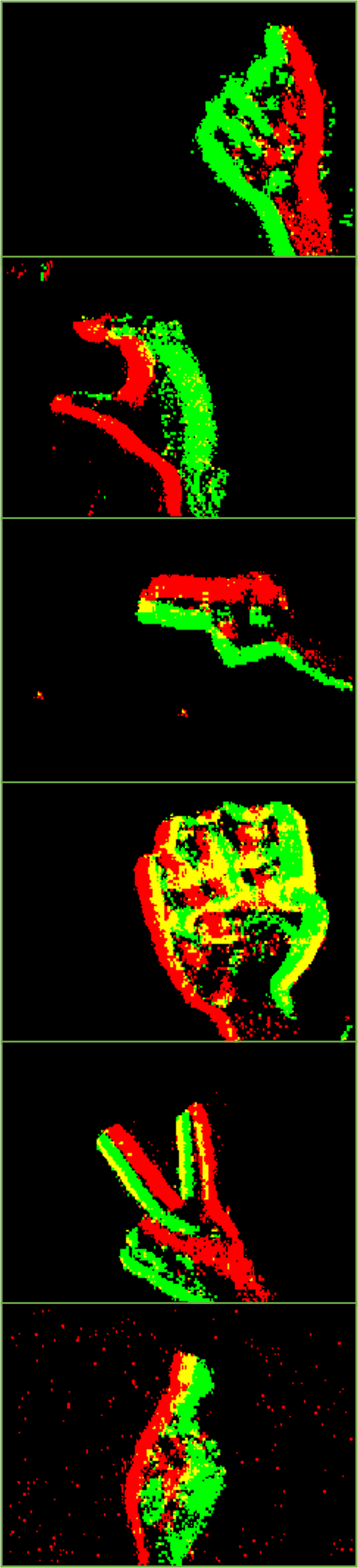}
		 \caption{Siqi Li et al. (SNN) \cite{snn}}\label{fig:5c}
	\end{subfigure}	
        \begin{subfigure}{0.17\linewidth}
		\includegraphics[width=\linewidth]{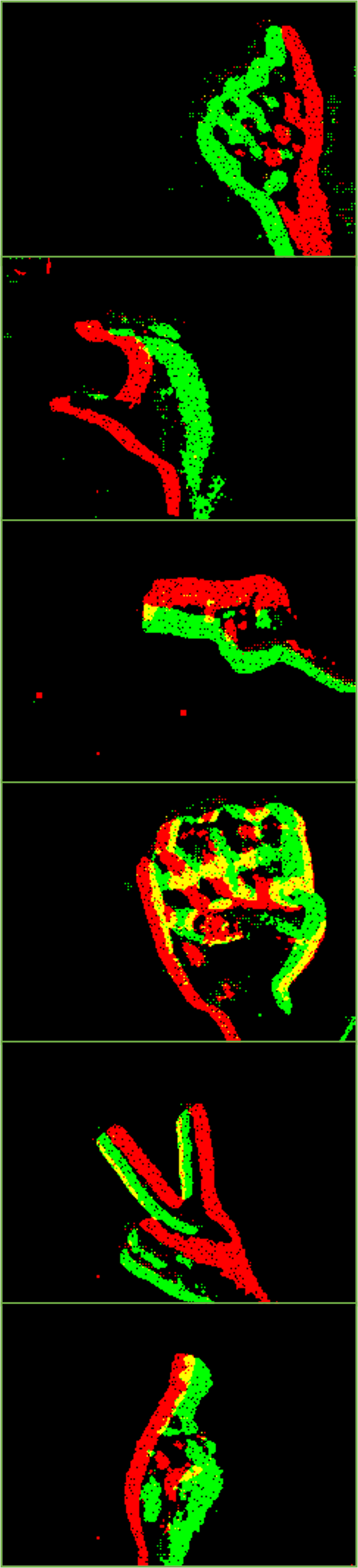}
		 \caption{Proposed SDNN}\label{fig:5d}
	  \end{subfigure}
	
 \caption{The qualitative analysis of the ASL-DVS dataset compares the performance of SNN \cite{snn} approach with the proposed Sigma Delta Neural Network. Each row presents reconstructed image of input-LR, ground truth-HR, predicted SNN - HR and the proposed SDNN predicted HR, allowing for a direct comparison. (Best viewed in 2x Zoom).}
 \label{fig:asl}
\end{figure}



\subsection{Qualitative Results}

\cref{fig:asl} shows the  ASL-DVS dataset qualitative performance exhibited by the proposed SDNN when compared to existing SNN models (and supplementary material contains results from NMNIST and CIFAR-DVS). Upon examining the figure, emphasizing the importance of clarity/precision in sign language, it becomes evident that the SDNN surpasses traditional SNN models by showcasing noise resilience and the ability to selectively filter events at lower temporal resolutions. Its design includes a significant threshold feature for temporal resolution, ensuring only relevant events exceeding this threshold are processed. This capability not only minimizes the impact of noise but also prioritizes critical events, enhancing its efficiency in super-resolution tasks by focusing on temporally significant information, thereby streamlining event-based data processing.

\subsection{Robustness to Noise}
To provide a more comprehensive evaluation, this sections includes quantitative metrics such as Peak Signal-to-Noise Ratio (PSNR), in addition to qualitative results. \cref{tab:noise} compares the PSNR of high-resolution (HR) streams against the predicted enhanced streams of SNN and the proposed SDNN. This will help establish noise performance benchmarks for our SDNN by comparing it with HR streams and traditional SNNs.

\begin{table}[h!]
    \caption{Quantitative Noise Evaluation: Comparison of PSNR (dB) for High Resolution (HR) vs SNN \cite{snn} and proposed SDNN.}
    \centering
    \begin{tabular}{lccc}
        \toprule
        \textbf{Dataset} & \textbf{HR vs. SNN \cite{snn}} & &\textbf{HR vs. Proposed SDNN} \\
        \midrule
        {N-MNIST}  & 20.60 & & \textbf{21.10} \\
        {CIFAR10-DVS}  & 21.28 & & \textbf{21.62} \\
        {ASL-DVS} &  32.74 & & \textbf{33.82} \\
        \bottomrule
    \end{tabular}
    \label{tab:noise}
\end{table}

\section{Computational Complexity}
\label{sec:cc}

The efficiency of neural network models is closely linked to their computational complexity. In this study, we examine this by training the N-MNIST dataset using the proposed Sigma Delta Neural Network (SDNN) and Spiking \& Artificial Neural Networks, all sharing the same network architecture. \cref{cc} compares SDNN, Spiking Neural Network (SNN), and Artificial Neural Network (ANN) based on key metrics such as the number of activations (or events) and sparsity.

Synaptic operations ("synops") refer to the computations associated with neural network connections, particularly in biologically inspired models, and represent the adjustments or use of synaptic weights during forward and backward passes. In contrast, "MACs" (Multiply-Accumulate Operations) quantify fundamental arithmetic operations, especially in convolutional and fully connected layers, and provide a general measure of the computational workload needed to compute the weighted sum of inputs. Both metrics are crucial for assessing the efficiency and computational complexity of neural network architectures. (The pseudo code for calculating computational complexity is outlined in Algorithm 2, provided in the supplementary material.)

\begin{table}[h]
\caption{Computational Complexity of SDNN, SNN, and ANN}
    \centering
    \begin{tabular*}{\textwidth}{@{\extracolsep{\fill}} l c r r r r r r}
        \toprule
        & & \multicolumn{2}{c}{\textbf{Proposed-SDNN}} & \multicolumn{2}{c}{\textbf{SNN}} & \multicolumn{2}{c}{\textbf{ANN}} \\
        \cmidrule(lr){3-4} \cmidrule(lr){5-6} \cmidrule(lr){7-8}
        Layers & Shape & \multicolumn{1}{c}{Events} & \multicolumn{1}{c}{Synops} & \multicolumn{1}{c}{Events} & \multicolumn{1}{c}{Synops} & \multicolumn{1}{c}{Activations} & \multicolumn{1}{c}{MACs} \\
        \toprule
        layer-0 & (17, 17, 2) & 15.43 & - & 25.74 & - & 578 & - \\
        layer-1 & (17, 17, 8) & 239.27 & 3085.75 & 393.89 & 5148.69 & 2312 & 115600 \\
        layer-2 & (34, 34, 2) & 151.66 & 478.54 & 732.04 & 787.79 & 2312 & 4624 \\
        layer-3 & (34, 34, 2) & 34.58 & 303.32 & 0.00 & 1464.08 & 2312 & 4624 \\
        \midrule
        \textbf{Total} & & \textbf{440.94} & \textbf{3867.61} & 1151.68 & 7400.56 & 7514 & 124848 \\
        \bottomrule
    \end{tabular*}
    \label{cc}
\end{table}


In this comparison, the ANN involves 7,514 activations and 124,848 MACs. In contrast, both SDNN and SNN exhibit significant sparsity in activations and synaptic operations (synops). SDNN, with a total of 440.94 activations, shows a substantial reduction in events compared to ANN, achieving 17.04 times better event sparsity and 32.28 times better synops sparsity. This indicates that SDNN enhances efficiency by minimizing both activations and synaptic operations.

Similarly, SNN has 1,151.68 activations, fewer than ANN but more than SDNN. SNN's event sparsity is 6.52 times, and synops sparsity is 16.87 times better than ANN, highlighting its computational efficiency through reduced synaptic operations and improved use of activations. Sparsity is particularly pronounced in SDNN's layer-3, where the event count drops significantly to 34.58, compared to no activity (0.00) in SNN, indicating highly sparse activation in deeper layers.

Unlike SDNN and SNN, the ANN lacks a spiking mechanism and operates with continuous activation functions, resulting in constant activations across layers, where each neuron is always 'active' to some extent. This contrasts sharply with spiking models, where neuron activity is event-driven and can be zero. Overall, both SDNN and SNN demonstrate better computational efficiency than traditional ANNs, as shown by their lower activations and synops, with SDNN having a slight edge in efficiency based on its sparsity metrics.
\section{Follow-on Implementation}
The neuromorphic-event field lacks super-resolution datasets with varying resolutions due to calibration and synchronization challenges between different event cameras. To address this, \cite{ref14} introduced the E-NFS event super-resolution dataset, the first of its kind, providing real-event data with LR-HR pairs for 2x and 4x scaling. Additionally, \cite{ref20} introduced 8x and 16x pairs, but only 2x and 4x are publicly available. These pairs were recorded using a display-camera setup, with high-resolution at 222 x 124 (4x), low-resolution at 111 x 64 (2x), and 56 x 31 (1x). We extended our network to include training on the E-NFS dataset, with results detailed in \cref{tab3}. The proposed SDNN model achieves RMSE scores of 0.515 for 2x and 0.546 for 4x, outperforming methods like Bi-cubic, SRFBN \cite{ref21}, EventZoom \cite{ref14}, and RNN \cite{ref20}. While the 4x results are noted, the study focuses on 2x super-resolution for detailed performance comparisons within a practical framework, optimizing computational efficiency and training stability.

 \begin{table}[h]
\centering
    \caption{Quantitative comparison of the proposed SDNN model against existing methods on the ENFS-real dataset, showing average spatial and temporal RMSE values for 2xSR and 4xSR upscaling factors.}
    \begin{tabular}{lcc}
        \toprule
        \textbf{Model} & \textbf{E-NFS-real} (2xSR) & \textbf{E-NFS-real} (4xSR) \\
        \midrule
        Bi-cubic \cite{ref20} & 0.899 & 0.969\\
        EventZoom \cite{ref14} &  0.773 & 0.910\\
        SRFBN \cite{ref21} &  0.669 & 0.753\\
        RNN \cite{ref20} & 0.663 & 0.663\\
        \midrule
        Proposed SDNN\textsuperscript{ST} &  \textbf{0.515} & \textbf{0.546}\\
        \midrule
    \end{tabular}
    \label{tab3}
\end{table}

\subsection{Downstream Application}
To evaluate the efficacy of the proposed approach and the quality of high-resolution (HR) event streams, this study uses object classification as a downstream application. The evaluation involves classifying low-resolution (LR) event streams, ground truth HR event streams, HR event streams generated by the proposed SDNN method, and those from \cite{li} and \cite{snn}. The classifier from Gehrig et al. \cite{gehrigICCV}, as adopted in \cite{snn}, is used for consistent comparison.
\begin{table}[h!]
    \caption{Object Classification Evaluation: Comparison of classification accuracy between the proposed SDNN Model and existing methods as reported in \cite{li} and \cite{snn} for Event Stream Super Resolution}
    \centering
    \begin{tabular}{lccc}
        \toprule
        \textbf{Model} & \textbf{N-MNIST} & \textbf{CIFAR10-DVS} & \textbf{ASL-DVS} \\
        \midrule
        \textbf{LR-Events}  & 90.0\% & 50.8\% & 99.7\% \\

        \textbf{li et al.} \cite{li} & 97.8\% & 59.6\% & 98.0\% \\

        \textbf{Siqi et al.} \cite{snn} & 99.1\% & \textbf{76.8\%} & \textbf{99.9\%} \\

        \midrule
        \textbf{Predicted HR SDNN} & \textbf{99.5\%} & {64.1\%} & {99.8\%} \\
             
        \textbf{HR- Events} \emph{(ref)}  & 99.1\% &  78.7\% &  99.9\%\\
        \bottomrule
    \end{tabular}
    \label{tab:class_results}
    
\end{table}

\cref{tab:class_results} compares the classification accuracy of the proposed SDNN model with methods by Li et al. \cite{li} and Siqi et al. \cite{snn} across N-MNIST, CIFAR10-DVS, and ASL-DVS datasets. On N-MNIST, SDNN leads with 99.5\% accuracy, outperforming LR events (90.0\%), Li et al. (97.8\%), and Siqi et al. (99.1\%). In CIFAR10-DVS, Siqi et al. achieve 76.8\%, HR events 78.7\%, and SDNN 64.1\%. For ASL-DVS, Siqi et al. and HR events reach 99.9\%, while SDNN achieves 99.8\%, surpassing LR events (99.7\%) and Li et al. (98.0\%). This shows SDNN's strong performance on N-MNIST and areas for improvement on CIFAR10-DVS and ASL-DVS.

\section{Ablation Study}
This study conducts ablation experiments using binary spikes as input into SDNN. \cref{tab:results} and \cref{tab3} presents the quantitative outcomes for proposed model. Beyond accuracy, SDNN's notable attribute is its resilience to noise, as illustrated in \cref{fig:asl} and \cref{tab:noise}, showing the model's robustness to noise in event data. Additionally, the computational efficiency of the proposed models, compared to similar architecture SNN and ANN, is critically analyzed in  \cref{sec:cc}. \cref{cc} illustrates a trend where the number of activations/events in SDNNs decreases as the network deepens, signifying efficiency in sparsity and MAC/synops. The experiment's algorithms are detailed in the supplementary materials.

\subsection{Limitations}
This study's modulation strategy involves integrating various parameters from the SDNN's neurons (as discussed in \cref{sec:network}, 4th paragraph), which poses challenges in achieving optimal parameter balance, also presents opportunities for further optimization. Additionally, future research focusing on real-time inferencing analysis on neuromorphic chips and embedded boards would be beneficial.

\section{Conclusion}
This paper introduces a novel end-to-end sigma-delta neural network (SDNN) architecture designed for efficient event stream processing, converting low resolution inputs into high resolution event data. This approach is validated against benchmark datasets such as NMNIST, CIFAR-DVS, ASL-DVS, and E-NFS, demonstrating state of the art performance in generating high-resolution event streams. Notably, the proposed SDNN significantly reduces computational complexity, outperforming traditional artificial neural networks (ANNs) and spiking neural networks (SNNs). The SDNN achieves 17.04 times better event sparsity and 32.28 times better synaptic operations sparsity compared to an equivalent ANN, and it outperforms SNNs by a factor of two. By focusing on the temporal differences and efficiently integrating these differences across time, SDNNs open new possibilities for the development of efficient neural network models.

\section*{Acknowledgement} This work was supported in part by the Irish Research Council (IRC) under Employment-Based Ph.D. Scheme with host institution as the University of Galway, Ireland, and industry mentor FotoNation Ltd., under Grant EBPPG/2022/17. The authors would also like to thank Dr Petronel Bigioi for providing the initial motivation and ideas that inspired us to work on this topic.

\bibliographystyle{splncs04}
\bibliography{main}

\begin{thebibliography}{10}
\providecommand{\url}[1]{\texttt{#1}}
\providecommand{\urlprefix}{URL }
\providecommand{\doi}[1]{https://doi.org/#1}

\bibitem{asl}
Bi, Y., Chadha, A., Abbas, A., , Bourtsoulatze, E., Andreopoulos, Y.: Graph-based object classification for neuromorphic vision sensing. In: 2019 IEEE International Conference on Computer Vision (ICCV). IEEE (2019)

\bibitem{ref5}
Courbariaux, M., Hubara, I., Soudry, D., El-Yaniv, R., Bengio, Y.: Binarized neural networks: Training deep neural networks with weights and activations constrained to+ 1 or-1. arXiv preprint arXiv:1602.02830  (2016)

\bibitem{ref4}
Diehl, P.U., Neil, D., Binas, J., Cook, M., Liu, S.C., Pfeiffer, M.: Fast-classifying, high-accuracy spiking deep networks through weight and threshold balancing. In: 2015 International Joint Conference on Neural Networks (IJCNN). pp.~1--8 (2015). \doi{10.1109/IJCNN.2015.7280696}

\bibitem{neurozoom}
Duan, P., Ma, Y., Zhou, X., Shi, X., Wang, Z.W., Huang, T., Shi, B.: Neurozoom: Denoising and super resolving neuromorphic events and spikes. IEEE Transactions on Pattern Analysis and Machine Intelligence  \textbf{45}(12),  15219--15232 (2023). \doi{10.1109/TPAMI.2023.3304486}

\bibitem{ref14}
Duan, P., Wang, Z.W., Zhou, X., Ma, Y., Shi, B.: Eventzoom: Learning to denoise and super resolve neuromorphic events. In: Proceedings of the IEEE/CVF Conference on Computer Vision and Pattern Recognition. pp. 12824--12833 (2021)

\bibitem{gehrigICCV}
Gehrig, D., Loquercio, A., Derpanis, K., Scaramuzza, D.: End-to-end learning of representations for asynchronous event-based data. In: 2019 IEEE/CVF International Conference on Computer Vision (ICCV). pp. 5632--5642 (2019). \doi{10.1109/ICCV.2019.00573}

\bibitem{ref16}
Guo, G., Feng, Y., Lv, H., Zhao, Y., Liu, H., Bi, G.: Event-guided image super-resolution reconstruction. Sensors  \textbf{23}(4), ~2155 (2023)

\bibitem{ref15}
Han, J., Yang, Y., Zhou, C., Xu, C., Shi, B.: Evintsr-net: Event guided multiple latent frames reconstruction and super-resolution. In: Proceedings of the IEEE/CVF International Conference on Computer Vision. pp. 4882--4891 (2021)

\bibitem{dvs}
IniVation: {DAVIS346 Event Camera}. \url{https://inivation.com/buy/} (2022), [Online; accessed 04-March-2024]

\bibitem{ref13}
Jing, Y., Yang, Y., Wang, X., Song, M., Tao, D.: Turning frequency to resolution: Video super-resolution via event cameras. In: Proceedings of the IEEE/CVF Conference on Computer Vision and Pattern Recognition. pp. 7772--7781 (2021)

\bibitem{lava}
Lava-DL: {lava-dl is a library of deep learning tools within Lava that support offline training, online training and inference methods for various Deep Event-Based Networks}. \url{https://github.com/lava-nc/lava-dl} (2022), [Online; accessed 27-November-2023]

\bibitem{li}
Li, H., Li, G., Shi, L.: Super-resolution of spatiotemporal event-stream image. Neurocomputing  \textbf{335},  206--214 (2019)

\bibitem{cifar}
Li, H., Liu, H., Ji, X., Li, G., Shi, L.: Cifar10-dvs: an event-stream dataset for object classification. Frontiers in neuroscience  \textbf{11}, ~309 (2017)

\bibitem{snn-git}
Li, S.: {Dataset contamination}. \url{https://github.com/lisiqi19971013/EventStream-SR/issues/7} (2021), [Online; accessed 27-November-2023]

\bibitem{ref21}
Li, Z., Yang, J., Liu, Z., Yang, X., Jeon, G., Wu, W.: Feedback network for image super-resolution. In: 2019 IEEE/CVF Conference on Computer Vision and Pattern Recognition (CVPR). pp. 3862--3871 (2019). \doi{10.1109/CVPR.2019.00399}

\bibitem{ref8}
Lichtsteiner, P., Posch, C., Delbruck, T.: A 128$\times$ 128 120 db 15 $\mu$s latency asynchronous temporal contrast vision sensor. IEEE Journal of Solid-State Circuits  \textbf{43}(2),  566--576 (2008). \doi{10.1109/JSSC.2007.914337}

\bibitem{ref18}
Lu, Y., Wang, Z., Liu, M., Wang, H., Wang, L.: Learning spatial-temporal implicit neural representations for event-guided video super-resolution. In: Proceedings of the IEEE/CVF Conference on Computer Vision and Pattern Recognition. pp. 1557--1567 (2023)

\bibitem{ref11}
Mostafavi, M., Nam, Y., Choi, J., Yoon, K.J.: E2sri: Learning to super-resolve intensity images from events. IEEE Transactions on Pattern Analysis and Machine Intelligence  \textbf{44}(10),  6890--6909 (2022). \doi{10.1109/TPAMI.2021.3096985}

\bibitem{ref10}
O'Connor, P., Gavves, E., Welling, M.: Temporally efficient deep learning with spikes. arXiv preprint arXiv:1706.04159  (2017)

\bibitem{ref9}
O'Connor, P., Welling, M.: Deep spiking networks. arXiv preprint arXiv:1602.08323  (2016)

\bibitem{sdnn}
O'Connor, P., Welling, M.: Sigma delta quantized networks. In: International Conference on Learning Representations (2016)

\bibitem{nmnsit}
Orchard, G., Jayawant, A., Cohen, G.K., Thakor, N.: Converting static image datasets to spiking neuromorphic datasets using saccades. Frontiers in neuroscience  \textbf{9}, ~437 (2015)

\bibitem{genx}
Prophesee: {Metavision - GenX320 Camera}. \url{https://www.prophesee.ai/event-based-sensor-genx320/} (2023), [Online; accessed 04-March-2024]

\bibitem{sekikawa2021learning}
Sekikawa, Y., Uto, K.: Learning to sparsify differences of synaptic signal for efficient event processing  (2021)

\bibitem{shariff}
Shariff, W., Dilmaghani, M.S., Kielty, P., Moustafa, M., Lemley, J., Corcoran, P.: Event cameras in automotive sensing: A review. IEEE Access  \textbf{12},  51275--51306 (2024). \doi{10.1109/ACCESS.2024.3386032}

\bibitem{snn}
Siqi, L., Feng, Y., Li, Y., Jiang, Y., Zou, C., Gao, Y.: Event stream super-resolution via spatiotemporal constraint learning. In: 2021 IEEE/CVF International Conference on Computer Vision (ICCV). pp. 4460--4469 (2021). \doi{10.1109/ICCV48922.2021.00444}

\bibitem{ref12}
Wang, L., Kim, T.K., Yoon, K.J.: Eventsr: From asynchronous events to image reconstruction, restoration, and super-resolution via end-to-end adversarial learning. In: Proceedings of the IEEE/CVF Conference on Computer Vision and Pattern Recognition. pp. 8315--8325 (2020)

\bibitem{ref20}
Weng, W., Zhang, Y., Xiong, Z.: Boosting event stream super-resolution with a recurrent neural network. In: European Conference on Computer Vision. pp. 470--488. Springer (2022)

\bibitem{ref19}
Xiang, X., Zhu, L., Li, J., Wang, Y., Huang, T., Tian, Y.: Learning super-resolution reconstruction for high temporal resolution spike stream. IEEE Transactions on Circuits and Systems for Video Technology  (2021)

\bibitem{ref3}
Yoon, Y.C.: Lif and simplified srm neurons encode signals into spikes via a form of asynchronous pulse sigma--delta modulation. IEEE transactions on neural networks and learning systems  \textbf{28}(5),  1192--1205 (2016)

\bibitem{ref17}
Yu, L., Wang, B., Zhang, X., Zhang, H., Yang, W., Liu, J., Xia, G.S.: Learning to super-resolve blurry images with events. IEEE Transactions on Pattern Analysis and Machine Intelligence  (2023)

\bibitem{ref2}
Zambrano, D., Bohte, S.M.: Fast and efficient asynchronous neural computation with adapting spiking neural networks. arXiv preprint arXiv:1609.02053  (2016)

\end{thebibliography}
\end{document}